\documentclass[superscriptaddress,twocolumn,showpacs,preprintnumbers,amsmath,amssymb]{revtex4}
\usepackage{graphicx}
\usepackage{dcolumn}
\usepackage{bm}
\usepackage{latexsym}
\usepackage[dvips]{color}
\begin{document}
\title{Footprint traversal by ATP-dependent chromatin remodeler motor}
\author{Ashok Garai} 
\affiliation{Department of Physics, Indian Institute of Technology,
Kanpur 208016, India} 
\author{Jesrael Mani} 
\affiliation{Department of Physics, Indian Institute of Technology,
Kanpur 208016, India} 
\author{Debashish Chowdhury{\footnote{Corresponding author(E-mail: debch@iitk.ac.in)}}} 
\affiliation{Department of Physics, Indian Institute of Technology,
Kanpur 208016, India} 
\affiliation{Max-Planck Institute for Physics of Complex Systems, 01187 Dresden, Germany}
\date{\today}
\begin{abstract}
ATP-dependent chromatin remodeling enzymes (CRE) are bio-molecular 
motors in eukaryotic cells. These are driven by a chemical fuel, 
namely, adenosine triphosphate (ATP). CREs actively participate 
in many cellular processes that require accessibility of specific 
segments of DNA which are packaged as chromatin. The basic unit 
of chromatin is a nucleosome where 146 bp $\sim$ 50 nm of a 
double stranded DNA (dsDNA) is wrapped around a spool formed by 
histone proteins. The helical path of histone-DNA contact on a  
nucleosome is also called ``footprint''. We investigate the mechanism 
of footprint traversal by a CRE that translocates along the dsDNA. 
Our two-state model of a CRE captures effectively two distinct 
chemical (or conformational) states in the mechano-chemical cycle 
of each ATP-dependent CRE. We calculate the mean time of traversal. 
Our predictions on the ATP-dependence of the mean traversal time  
can be tested by carrying out {\it in-vitro} experiments on 
mono-nucleosomes. 
\end{abstract}
\pacs{87.16.ad,87.16.Sr,87.16.dj}
\maketitle
\vspace{2mm}
\section{Introduction}

A deoxyribonucleic acid (DNA) molecule is a linear heteropolymer whose 
monomeric subunits, called nucleotides, are denoted by the four letters 
A, T, C and G. The sequence of these nucleotides in a DNA molecule 
chemically encodes genetic informations. In the nucleus of an eukaryotic 
cell, DNA is stored in a hierarchically organized structure called 
chromatin \cite{wolffe98,widom98,schiessel03,lavelle06}. 
The primary repeating unit of chromatin at the lowest level of the 
hierarchical structure is a nucleosome \cite{kornberg99}. 
The cylindrically shaped core of each nucleosome consists of an octamer 
of histone proteins around which 146 base pairs (i.e.,$\sim 50$ nm) of the 
the double stranded DNA is wrapped about two turns (more precisely, 1.7 
helical turns); the arrangement is reminiscent of wrapping of a thread 
around a spool. There are $14$ equispaced sites, at intervals of $10$ 
base pairs (bp), on the surface of the cylindrical spool. Electrostatic 
attraction between these binding sites on the histone spool and the 
oppositely charged DNA seems to dominate the histone-DNA interactions 
which stabilize the nucleosomes. Throughout this paper, the helical 
curve formed by the histone-DNA overlap will be called the ``footprint''. 

The DNA stores the genetic blueprint of an organism. If nucleosomes 
were static, segments of DNA buried in nucleosomes would not be  
accessible for various functions involving the corresponding gene 
\cite{polach95,anderson02,fan03}. 
However, in reality, nucleosomes are dynamic. Spontaneous dynamics  
of nucleosomes are usually consequences of thermal fluctuations whereas 
the active dynamic processes are driven by special purpose molecular 
machines called chromatin remodeling enzymes (CRE) fuelled by ATP 
\cite{flaus03,flaus11,halford04a,saha06,clapier09,narlikar08}. 
Various aspects of chromatin dynamics has received some attention of  
theoretical modelers, including physicists, over the last few years 
\cite{sakaue01,schiessel01,kulic03a,kulic03b,schiessel06,rafiee04,blossey11,mobius06,langowski06,lense06,vaillant06,ranjith07,lia06}.

In general, ``Chromatin remodelling'' refers to a range of enzyme 
mediated structural transitions that occur during gene regulation 
in eukaryotic cells. To make DNA, which is wrapped around histone 
octamer, accessible for various DNA-dependent processes, it is 
always necessary to rearrange or mobilize the nucleosomes. 
In principle, there are at least four different ways in which a CRE 
can affect the nucleosomes \cite{cairns07}:  
(i) {\it sliding} the histone octamer, i.e., repositioning of the
entire histone spool, on the dsDNA; (ii) {\it exchange} of one 
or more of the histone subunits of the spool with those in the
surrounding solution (also called {\it replacement} of histones)
(iii) {\it removal} of one or more of the histone subunits of the
spool, leaving the remaining subunits intact, and (iv) complete
{\it ejection} of the whole histone octamer without replacement.
Our theoretical work here is closely related to {\it sliding}. 

In the next section we describe a scenario in which either a CRE 
motor (or, other ATP-dependent motors that translocate along dsDNA) 
traverse the ``footprint''. Because of the stochasticity of the 
underlying mechano-chemical kinetics, the {\it footprint traversal time} 
(FTT) is a fluctuating random variable. Extending an earlier model 
developed by Chou \cite{chou}, we analytically calculate the {\it 
mean} FTT (MFTT) of the CRE motor.

To our surprise, we found that the ATP-dependence of the various 
ATP-driven activities of CREs have not been studied systematically 
in the published literature. In particular, we address the question: 
how does the MFTT of a CRE motor vary 
with the variation of the concentration of ATP? This rate is not 
necessarily directly proportional to the rate of ATP hydrolysis by 
the CRE because the mechanical sliding of the nucleosome need not be 
tightly coupled with the  hydrolysis of ATP by the CRE. 
Therefore, we develope here an analytical theory predicting the 
ATP-dependence of the ATP-dependent footprint-traversal by CRE. 
We hope our result will stimulate systematic experimental investigations 
on the ATP-dependence of ATP-dependent CREs.

\section{CRE: phenomenology and motivation for this work}

Chromatin is not a frozen static aggregate of DNA and proteins.
Spontaneous thermal fluctuations can cause a transient unwrapping and
rewrapping of the nucleosomal DNA from one end of the nucleosome spool;
the corresponding rates for an isolated single nucleosome are, typically,
$4$s$^{-1}$ and $20-90$s$^{-1}$, respectively \cite{li04}.
In other words, once wrapped fully, the nucleosomal DNA remains in that
state for about $250$ms before unwrapping again spontaneously; however,
it waits in the unwrapped state only for about $10-50$ms before
rewrapping again spontaneously. Surprisingly, the accessibility of the
nucleosomal DNA is only modestly affected if instead of a single
nucleosome the experiment is repeated with an array of homogeneously
distributed nucleosomes \cite{poirier08,poirier09}.
Moreover, folding of an array of nucleosomes makes the linker DNA
about $50$ times less accessible \cite{poirier08,poirier09}.
Furthermore, on the nucleosomal DNA, the farther is a site from the
entry and exit points, the longer one has to wait to access it by
the rarer spontaneous fluctuation of sufficiently large size
\cite{tims11}.
Thus, nucleosomal DNA far from both the entry and exit sites is
practically inaccessible by spontaneous thermal fluctuations.

Can a nucleosome slide {\it spontaneously} by thermal fluctuations 
thereby exposing the nucleosomal DNA? Interestingly, spontaneous 
repositioning of nucleosome on DNA strands is a well known phenomenon
\cite{meerssman92}. 
How can one reconcile accessibility of nucleosomal DNA by such 
repositioning \cite{meerssman92} with the difficulty of access by 
unwrapping from either end of the spool \cite{tims11}?  
If the DNA were to move unidirectionally along its own superhelical 
contour on the surface of the histone, at every step it would have 
to first {\it transiently} detach simultaneously from all the $14$ 
binding sites and then reattach at the same sites after its contour 
gets shifted by $10$ bp (or multiples of $10$ bp). But, the energy 
cost of the simultaneous detachment of the DNA from all the $14$ 
binding sites is prohibitively large because the total energy of 
binding at the $14$ sites is about $75 k_BT$ 
\cite{blossey11,kulic03a,kulic03b}. 
 
But, why can't the cylindrical spool simply roll on the wrapped 
nucleosomal DNA thereby repositioning itself? If the nucleosome 
rolls by detaching DNA from one end of the spool, cannot it compensate 
this loss of binding energy by simulataneous attachment with a binding 
at the other end? If such an energy compensation were possible,  
detachment from only one binding site would be required at a time.   
But, the cylindrical spool has a finite size on which only a finite 
number ($14$) of binding sites for DNA are accomodated. Therefore, 
by rolling over the DNA the spool would not offer any vacant binding 
site to the DNA with which it can bind. This rolling mechanism would 
successfully lead to spontaneous sliding of the nucleosome only if 
the histone spool were infinite with an infinite sequence of binding 
sites for DNA on its surface \cite{blossey11}.

We now describe a plausible mechanism for spontaneous sliding of a 
nucleosome \cite{schiessel01,kulic03a,kulic03b}.
In the process of normal ``breathing'', most often the spontaneously 
unwrapped flap rewraps exactly to its original position on the 
histone surface. However, if the rewrapping of a unwrapped flap takes 
place at a slightly displaced location on the histone spool a small 
bulge (or loop) of DNA forms on the surface of the histones. Since 
the successive binding sites are separated by $10$bp, the length 
of the loop is quantized in the multiples of $10$bp \cite{schiessel01}.
Such a spontaneously created DNA loop, can diffuse in an unbiased 
manner on the surface of the histone spool. In the beginning of each 
step DNA from one end of the loop detaches from the histone spool, 
but the consequent energy loss is made up by the attachment of DNA 
at the other end of the loop to the histone spool before the step 
is completed. Consequently, by this diffusive dynamics, the DNA loop 
can traverse the entire length of the $14$ binding sites on the 
histone spool of a nucleosome which will manifest as sliding of the 
nucleosome by a length that is exactly equal to the length of DNA in 
the loop. The diffusing DNA bulge can be formed by a ``twist'', 
rather than bending, of DNA  \cite{lusser03,becker02a,langst04}. 
Spontaneous sliding of a nucleosome, however, is too slow to support 
intranuclear processes which need access to nucleosomal DNA. 

It is now widely agreed that ATP-dependent active remodeling of 
nucleosome can account for the fast sliding of nucleosomes.  
Nevertheless, bulging DNA loop is expected to play a key role 
in the remodeling process \cite{lia06}. 
Our model describes how a CRE motor can wedge itself at the fork between 
the histone spool and a transiently detached segment of dsDNA and, 
by exploiting the spontaneously diffusing loop by an ATP-dependent 
ratcheting, traverse the footprint in a directed manner. Because 
of the intrinsic stochasticity of the mechano-chemistry of the 
CRE and that of the diffusive motion of the DNA loop, the overall 
motion of the CRE is noisy and the time it takes to traverse the 
footprint is random.

The main question we address in this paper is the following: 
if $<T>$ is the MFTT, 
what is the dependence of $<T>$ on the concentration 
of ATP in the surrounding aqueous medium? 
To our knowledge, in the published literature neither systematic 
experiments nor any analytical theory has addressed this question. 
In this paper, by extending Chou's model \cite{chou} of CRE, we 
capture the role of ATP explicitly and derive an analytical 
expression for the dependence of $<T>$ on the concentration of ATP.  

A CRE may be regarded as a molecular motor where input energy 
is derived from ATP hydrolysis and the output is mechanical work. 
The directed movement of the CRE may be caused either by a {\it 
power stroke} or by a {\it Brownian ratchet} mechanism \cite{widom98}. 
The kinetics of the ATP-dependent CRE motor is formulated in our 
model in terms of a set of master equations. The kinetic scheme 
can be interpreted in terms of both power stroke and Brownian ratchet 
mechanisms.  

One interesting question \cite{cairns07} in the context of CRE is 
whether the CRE translocates along the DNA by moving around the 
nucleosome, or whether the CRE anchors on the histone octamer and 
``pumps'' DNA by pulling around the octamer. From the perspective 
of physicists, these two alternative scenarios can be viewed as 
merely a change of frame of reference- one is fixed with respect to 
the DNA whereas the other is fixed with respect to the CRE. Therefore, 
we describe the operation of the CRE with respect to a reference 
frame with respect to which the CRE translocates along the DNA; 
but, the model can be reformulated by a coordinate transformation so 
as to capture the alternative scenario where the CRE pumps the DNA.

\section{The Model}

\begin{figure}[ht]
\begin{center}
\includegraphics[width=0.85\columnwidth]{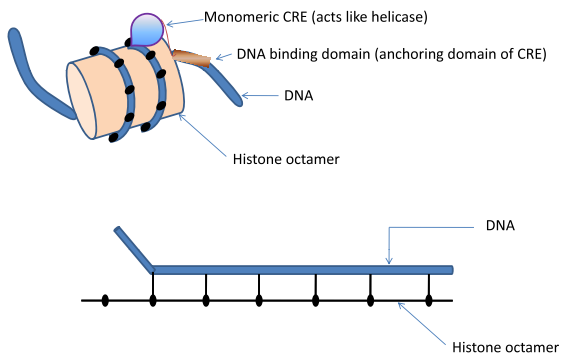}
\end{center}
\caption{A schematic representation of the isolated nucleosome.
}
\label{fig-spool}
\end{figure}

We model a mono nucleosome where a dsDNA is wrapped 
one-and-three-fourth turn around a disc-shaped spool made of 
histone proteins (see Fig.\ref{fig-spool}(a)). 
Following Chou \cite{chou}, we consider the scenario where 
the CRE ``wedges itself underneath the histone''. 

The sites of histone-DNA contact along the DNA chain is represented 
as a one-dimensional lattice. Therefore, the lattice constant is, 
typically, $10$ bp ((see Fig.\ref{fig-spool}(b)). The total 
number $n$ of lattice sites is equal to the total number of 
histone-DNA contact in a single nucleosome.

\subsection{Flap, loop and diffusive sliding of histone spool}

In this subsection we present a summary of Chou's ideas \cite{chou} 
which we need in the next subsection where we extend Chou's model. 
Here we consider the simple situation when no CRE is present and the 
kinetics of the system is governed solely by spontaneous thermal 
fluctuations. Because of these fluctuations, from either end of the 
histone-DNA contact region, small segments of DNA momentarily unwrap 
from the histone spool at a rate $k_u$. For energetic reasons, the most 
likely length of such a segment would be one lattice spacing, i.e., 
about $10$bp. Following Chou \cite{chou}, we call such unwrapped 
segments a ``flap''. The rate of the reverse transition, in which 
re-binding of the DNA flap with the histone, takes place at a rate $k_b$. 
 
A flap need not re-make the original histone-DNA contact. Instead, 
by pulling in an extra segment of the DNA, its next segment can 
bind with the last binding site on the histone spool, with rate 
$\alpha$ thereby forming what Chou \cite{chou} referred to as a 
``loop'. While located at either end of the lattice, a loop can 
revert to a flap at a rate $\beta$. The rates $\alpha$ and $\beta$ 
are well approximated by \cite{chou} 
\begin{equation} 
\alpha \sim k_b e^{-E_{bend}/(k_BT)}, ~{\rm and}~   \beta \sim k_u 
\label{eq-al_est1}
\end{equation} 
where $E_{bend}$ is the energy cost of bending the DNA into the 
shape of the loop.

\begin{figure}
\begin{center}
\includegraphics[scale=0.8]{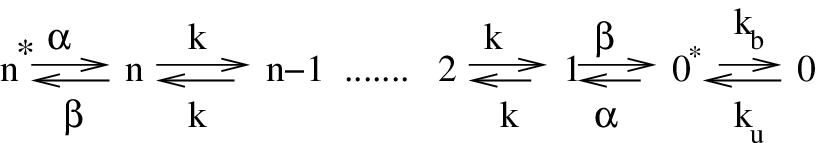}
\caption{A schematic representation of the position of the thermally generated flap and it's diffusion along wrapped DNA.}
\label{fig-diffn1}
\end{center}
\end{figure}

A loop can step  forward or backward. In the {\it absence} of any 
CRE, the rates of the forward and backward steppings of the loop 
are equal (denoted by $k$), provided the size of the loop $L_{loop}$ 
remains unaltered (see Fig.\ref{fig-diffn1}); in each forward step 
it unwraps one segment of DNA from the histone in the direction of 
its hop and re-wraps another equally long segment behind it. 
Therefore, one can approximate $k$ by \cite{chou}
\begin{equation}
k \sim k_{u}\biggl(\frac{k_b}{(k_b+k_u)}\biggr) 
\end{equation} 

When a loop, after entering the lattice from one end, makes an 
eventual exit from the other end, it completes the ``sliding'' 
of the histone spool by a distance $L_{loop}$ along the DNA in the 
opposite direction. Therefore, from the perspective of the sliding 
histone spool, its effective rate of hopping by a step of size 
$L_{loop}$ along the dsDNA strand is the same as the rate $p_{n}$ 
at which a DNA loop of length $L_{loop}$ traverses the lattice of 
$n$ sites from one end to the other. 

Suppose ${\cal P}_{j}(t)$ denotes the probability that the loop 
is located at $j$ ($0 \leq j \leq 1$).  Following Chou's arguments, 
based on master equations for ${\cal P}_{j}(t)$, one gets \cite{chou}
\begin{eqnarray}
p_{n}&=&\frac{\alpha kk_{u}}{(n-1)\beta k_{b}+k(\alpha+2k_{b})}
\end{eqnarray}
In the absence of a CRE, the traversal of a DNA loop of length 
$L_{loop}$ from left to right is as likely as that from right to 
left. Therefore, the histone spool can slide forward or backward, 
with equal rate $p_n$, by a step of size $L_{loop}$. As we'll 
see in the next subsection, peeling off of the DNA from the 
histone spool by a CRE motor keeps decreasing the effective 
value of $n$ which, in turn, increases the effective sliding rate 
$p_{n}$.

\subsection{Kinetics of CRE-driven directed sliding of histone spool}

Next, we consider the effect of DNA loop diffusion on the ATP-dependent 
translocation kinetics of a CRE. The model and results presented in this 
subsection are extensions of Chou's work \cite{chou} by incorporating 
explicitly a Brownian ratchet mechanism for CRE motors. 

As in ref.\cite{chou}, we assume that the step size of the CRE motor 
is identical to the length of the thermally generated DNA loop. 
Therefore, the mechanical movements of the CRE motor can be described 
as that of a ``particle'' on the one-dimensional lattice on which 
the equi-spaced sites denote the histone-DNA contact points. We denote 
the position of the CRE motor on this lattice by the integers $m$.
We now extend Chou's model \cite{chou} by exploiting a superficial 
similarity with the Garai-Chowdhury-Betterton (GCB) model \cite{garai} 
for the Brownian ratchet mechanism of monomeric helicase motors.

A DNA helicase unwinds a dsDNA and translocates along one of two 
strands. At any arbitrary instant of time, the configuration of the 
system looks very similar to that shown in Fig.\ref{fig-spool}(b) 
except that the surface of the DNA spool and the dsDNA would be 
replaced by the two strands of the dsDNA itself. The lattice constant 
is $1$bp in the case of a helicase whereas it is about $10$ bp in 
Fig.\ref{fig-spool}(b). In the Brownian ratchet mechanism, momentary 
local unwinding of a segment, typically, $1$ bp long, takes place 
at the fork by spontaneous thermal fluctuation; the opportunistic 
advance of the helicase merely prevents closure of the segment. 
Similarly, in the Brownian ratchet mechanism of the CRE, the CRE 
is assumed to ``wedge'' itself just in front of the DNA-histone fork. 
The CRE motor can move forward only if the segment in front of it 
is unwrapped by thermal fluctuation.

\begin{figure}[ht]
\begin{center}
\vspace{2cm}
\includegraphics[width=0.75\columnwidth]{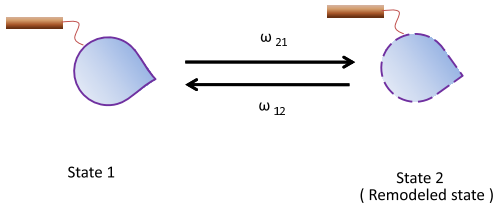}
\end{center}
\caption{A schematic representation of the transition of the CRE 
between its two states.
}
\label{fig-omega1221}
\end{figure}

\begin{figure}
\begin{center}
\includegraphics[width=0.75\columnwidth]{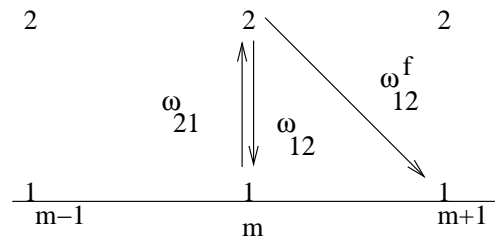}
\end{center}
\caption{A schematic representation of the position of the motor with two state of the model.}
\label{fig-model1}
\end{figure}

The mechano-chemical cycle of the CRE is captured in our model 
exactly the same way in which that of the helicase was formulated 
in the GCB model \cite{garai}.
We assume that sequence of states in each mechano-chemical cycle of 
a CRE can be combined into two distinct groups which we label by 
the integers $1$ and $2$ (see Fig.\ref{fig-omega1221}). The allowed 
transitions and the corresponding rate constants are shown in 
Fig.\ref{fig-model1}.

\begin{figure}
\begin{center}
\includegraphics[width=0.75\columnwidth]{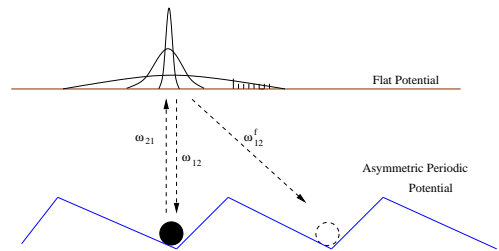}
\caption{A schematic representation of the Brownian ratchet mechanism.}
\label{fig-brrat}
\end{center}
\end{figure}

The physical processes captured by these rate constants can be motivated 
by a comparison with the abstract Brownian ratchet mechanism, illustrated 
in Fig.{\ref{fig-brrat}}. ATP hydrolysis by the CRE drives its transition 
from the state $1$ to the state $2$ at a rate $\omega_{21}$. Let us assume 
that the motor experiences two different types of potentials in the states 
$1$ and $2$. Let us further assume that initially the periodic potential, with 
asymmetric sawtooth-like period, is kept on for sometime and during this 
time the motor settles at a position that coincides with one of the minima 
of this potential. Now if this potential is switched off then the probability 
distribution of the position of the motor will spread as a symmetric Gaussian. 
After sometime this Gaussian profile is broad enough to overlap with the 
next well (shadded region in the Fig.{\ref{fig-brrat}}), in addition to the 
original well. Now if the sawtooth potential is again switched on then, 
with a non-zero probability (that is proportional to the area of the shaded 
region) the motor will find itself in the next well. Our model accounts for 
this possibility with the transition associated with the rate constant 
$\omega^{f}_{12}$. There is also a finite probability that the particle stays 
back in its original well; this is captured by the transition with the rate 
constant $\omega_{12}$. 

The CRE motor would step forward at the rate $\omega^f_{12}$ 
if the next site in front is cleared. But if the next site 
is not cleared and it has to wait for the unwrapping of the 
DNA segment by thermal fluctuation. Consequently, its {\it 
effective} hoping rate 
\begin{equation} 
\tilde\omega^{f}_{12}=k_{u}\biggl(\frac{\omega^{f}_{12}}{(\omega^{f}_{12}+k_{b})}\biggr) 
\label{eq-effwf12} 
\end{equation}
is reduced from the free hopping rate $\omega^{f}_{12}$ by a 
factor that depends on both $k_{u}$ and $k_{b}$. 

\begin{figure}
\begin{center}
(a)\\
\vspace*{10mm}
\includegraphics[scale=0.35]{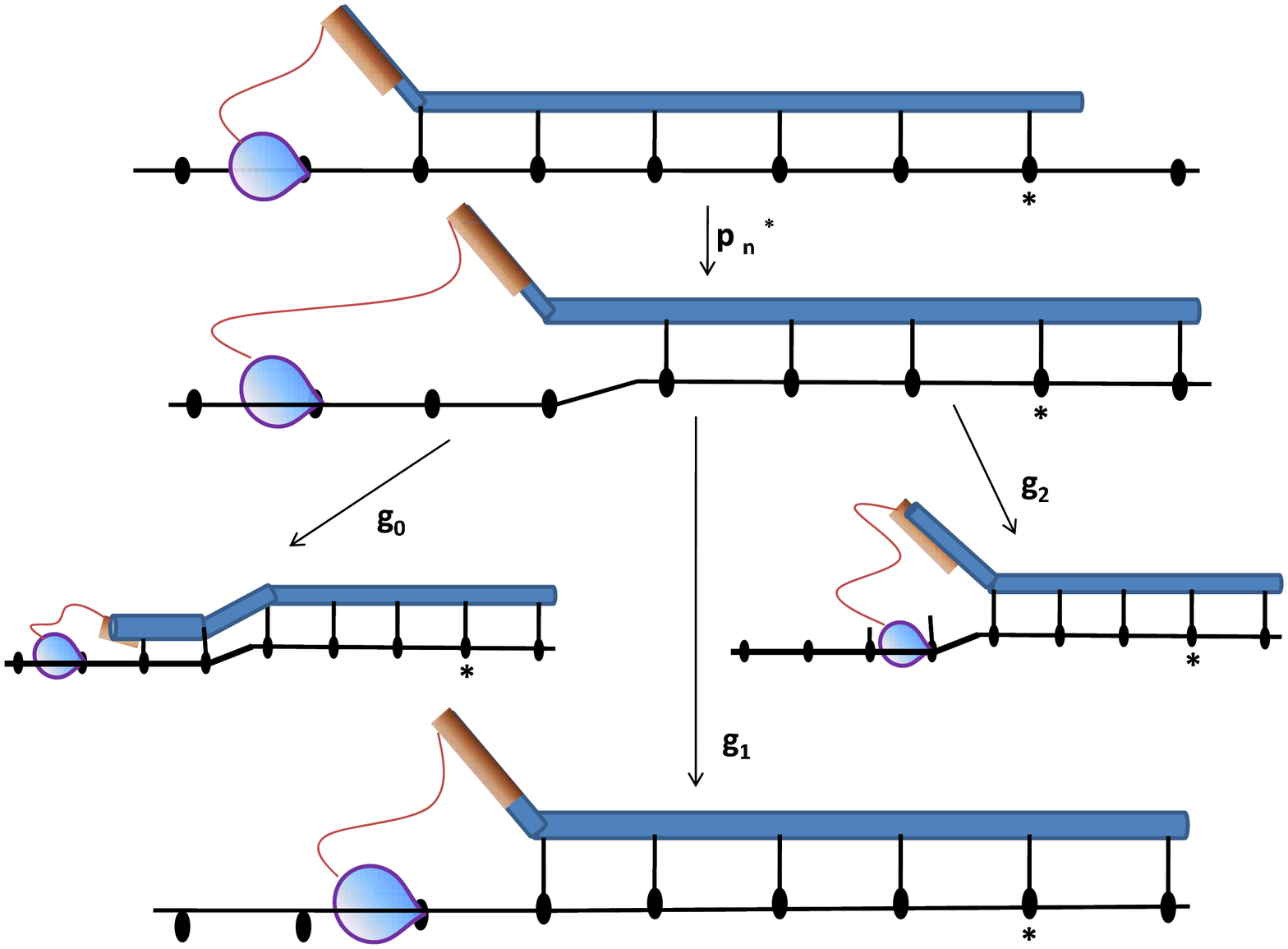} \\
\vspace*{10mm}
(b) \\
\vspace*{10mm}
\includegraphics[scale=0.35]{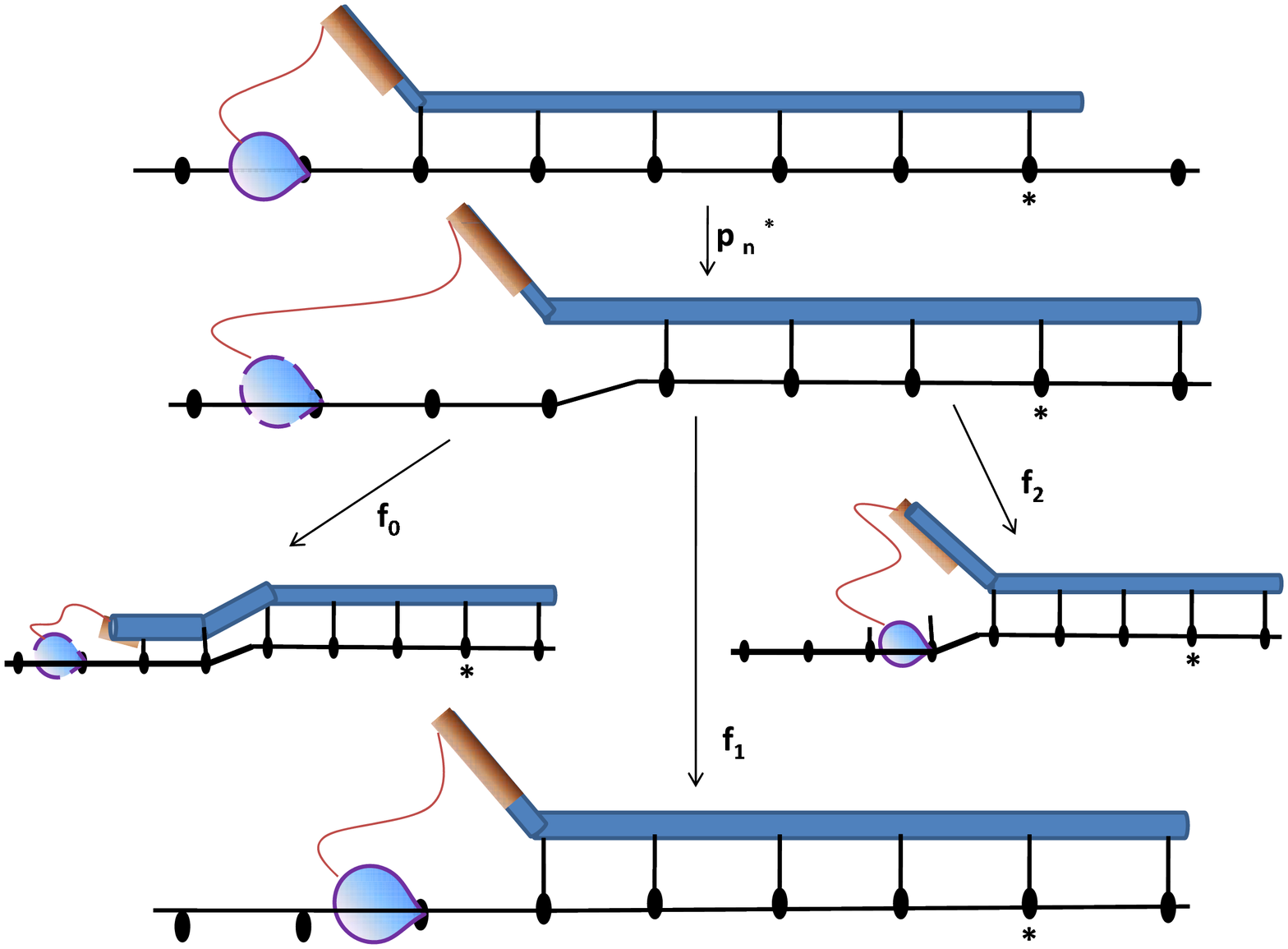}
\caption{Schematic representations of the possible transitions when 
the motor is in front of the open flap. In (a) and (b) the motor is 
in the chemical state $1$ and $2$, respectively.}
\label{fig-fate}
\end{center}
\end{figure}

When a diffusing loop reaches in front of the motor it momentarily 
creates a flap of two bond segments. Three different transitions 
are now possible (see Fig.\ref{fig-fate}): (i) the motor's position 
remains unaltered while the two open segments close, (ii) the motor 
moves forward by one step while one segment of the flap closes; 
(iii) the motor moves forward by two steps and the flap cannot close. 
The rate for the process (i) is $k_b/2$ irrespective of the ``chemical'' 
state of the motor. However, the rates of the processes (ii) and (iii) 
depend on whether the motor was in the ``chemical'' state $1$ or 
$2$. If the motor is in the state $2$, the rate of the process 
(ii) is given by $((\omega^{f}_{12})^{-1}+k_{b}^{-1})^{-1}$ 
and that of the process (iii) is given by 
$((\omega^{f}_{12})^{-1} + (\omega_{21})^{-1} + (\omega^{f}_{12})^{-1})$.
Therefore,
\begin{eqnarray}
f_{0}&=&\frac{k_{b}}{2\lambda_{f}},
f_{1}=\frac{\omega^{f}_{12}k_{b}}{(\omega^{f}_{12}+k_{b})\lambda_{f}},
f_{2}=\frac{\omega^{f}_{12}\omega_{21}}{\lambda_{f}(2\omega_{21}+\omega^{f}_{12})},
\end{eqnarray}
with the normalization constant 
\begin{eqnarray}
\lambda_{f}=\frac{k_{b}}{2}+\frac{\omega^{f}_{12}k_{b}}{(\omega^{f}_{12}+k_{b})}+\frac{\omega^{f}_{12}\omega_{21}}{(2\omega_{21}+\omega^{f}_{12})}
\end{eqnarray}
where the symbols $f_0$, $f_1$ and $f_2$ are the probabilities 
of the processes (i), (ii) and (iii) above when the motor is in 
the ``chemical'' state $2$. 
Similarly,
\begin{eqnarray}
g_{0}&=&\frac{k_{b}}{2\lambda_{h}},
g_{1}=\frac{k_{b}\omega_{21}\omega^{f}_{12}}{(\omega_{21}\omega^{f}_{12}+k_{b}\omega^{f}_{12}+k_{b}\omega_{21})\lambda_{h}},\nonumber\\
g_{2}&=&\frac{\omega_{21}\omega^{f}_{12}}{2(\omega^{f}_{12}+\omega_{21})\lambda_{h}}, 
\end{eqnarray}
with the normalization constant
\begin{eqnarray}
\lambda_{h}&=&\frac{k_{b}}{2}+\frac{k_{b}\omega_{21}\omega^{f}_{12}}{\omega_{21}\omega^{f}_{12}+k_{b}\omega^{f}_{12}+k_{b}\omega_{21}}\nonumber\\
&+&\frac{\omega_{21}\omega^{f}_{12}}{2(\omega^{f}_{12}+\omega_{21})}, 
\end{eqnarray}
are the corresponding probabilities, when the motor is in the 
``chemical'' state $1$. 

Suppose, $N$ is the maximum number of histone-DNA contacts possible in 
the nucleosome. Let $m$ denote the instantaneous position of the motor.  
$n$ is the distance between the motor and the far end of histone-DNA 
contact. The master equations for the probabilities $P(m,n,t)$ are as follows: 

For n$\geq$N+1
\begin{eqnarray}
dP_{1}(m,n)/dt &=& \omega_{12} P_{2}(m,n) - \omega_{21} P_{1}(m,n) \nonumber \\
&+& \omega^{f}_{12} P_{2}(m-1,n+1) \nonumber \\
&+& p_{N} [P_{1}(m,n+1) + P_{1}(m,n-1) \nonumber \\ 
&& ~~~~- 2 P_{1}(m,n)]  
\label{eq-mn1}
\end{eqnarray}
and
\begin{eqnarray}
dP_{2}(m,n)/dt &=& \omega_{21} P_{1}(m,n) - \omega_{12} P_{2}(m,n) \nonumber \\
&-& \omega^{f}_{12} P_{2}(m,n) \nonumber \\
&+& p_{N} [P_{2}(m,n+1)+P_{2}(m,n-1) \nonumber \\
&& ~~~~- 2 P_{2}(m,n)] 
\label{eq-mn2}
\end{eqnarray}
For n=N
\begin{eqnarray}
dP_{1}(m,N)/dt &=& \omega_{12} P_{2}(m,N) - \omega_{21} P_{1}(m,N) \nonumber \\
&+& \omega^{f}_{12} P_{2}(m-1,N+1) \nonumber \\ 
&+& p_{N} [P_{1}(m,N+1) - P_{1}(m,N)] \nonumber \\ 
&+& p_{N} [f_{1} P_{2}(m-1,N) \nonumber \\
&& ~~~~ + g_{1} P_{1}(m-1,N)] \nonumber \\  
&+& p_{N-1} g_{0} P_{1}(m,N-1) 
\label{eq-mn3}
\end{eqnarray}
and
\begin{eqnarray}
dP_{2}(m,N)/dt &=& \omega_{21}P_{1}(m,N) - \omega_{12}P_{2}(m,N) \nonumber \\  
&-& \tilde\omega^{f}_{12} P_{2}(m,N) \nonumber \\
&+& p_{N}[P_{2}(m,N+1) - P_{2}(m,N)] \nonumber \\
&+& f_{0}p_{N-1}P_{2}(m,N-1)
\label{eq-mn4}
\end{eqnarray}
For 3$\leq$ n $\leq$ N
\begin{eqnarray}
dP_{1}(m,n)/dt &=&  \omega_{12}P_{2}(m,n) - \omega_{21}P_{1}(m,n) \nonumber \\ 
&+& \tilde\omega^{f}_{12}P_{2}(m-1,n+1) - p_{n}P_{1}(m,n) \nonumber \\
&+&p_{n}[f_{1} P_{2}(m-1,n) + g_{1} P_{1}(m-1,n)] \nonumber \\
&+&p_{n+1}[f_{2} P_{2}(m-2,n+1) \nonumber \\
&& ~~~ + g_{2} P_{1}(m-2,n+1)] \nonumber \\
&+& g_{0}p_{n-1}P_{1}(m,n-1) 
\label{eq-mn5}
\end{eqnarray}
and
\begin{eqnarray}
dP_{2}(m,n)/dt&=& \omega_{21}P_{1}(m,n) -\omega_{12}P_{2}(m,n) \nonumber \\
&-& (\tilde\omega^{f}_{12} + p_{N}) P_{2}(m,n) \nonumber \\
&+& f_{0} p_{n-1}P_{2}(m,n-1) 
\label{eq-mn6}
\end{eqnarray}
For n$=$2
\begin{eqnarray}
dP_{1}(m,2)/dt &=& \omega_{12}P_{2}(m,2)-\omega_{21}P_{1}(m,2) \nonumber \\
&+&\tilde\omega^{f}_{12}P_{2}(m-1,3) - p_{2}P_{1}(m,2) \nonumber \\
&+& p_{2} [g_{1} P_{1}(m-1,2) + f_{1} P_{2}(m-1,2)] \nonumber \\
&+& p_{3} [g_{2} P_{1}(m-2,3) + f_{2} P_{2}(m-2,3)] \nonumber \\
\label{eq-mn7}
\end{eqnarray}
and
\begin{eqnarray}
dP_{2}(m,2)/dt&=& \omega_{21}P_{1}(m,2) - \omega_{12}P_{2}(m,2) \nonumber \\
&-& \tilde\omega^{f}_{12} P_{2}(m,2) - p_{2} P_{2}(m,2) \nonumber \\
\label{eq-mn8}
\end{eqnarray}
For n$=$1
\begin{eqnarray}
dP_{1}(m,1)/dt&=& \omega_{12}P_{2}(m,1) - \omega_{21}P_{1}(m,1) \nonumber \\
&+& \tilde\omega^{f}_{12}P_{2}(m-1,2) - k_{u}P_{1}(m,1) \nonumber \\
&+& p_{2} [f_{2} P_{2}(m-2,2) + g_{2} P_{1}(m-2,2)]  \nonumber \\
\label{eq-mn9}
\end{eqnarray}
and
\begin{eqnarray}
dP_{2}(m,1)/dt= \omega_{21}P_{1}(m,1)-\omega_{12}P_{2}(m,1) - k_{u}P_{2}(m,1) \nonumber \\
\label{eq-mn10}
\end{eqnarray}
\begin{figure}
\begin{center}
\includegraphics[scale=0.7]{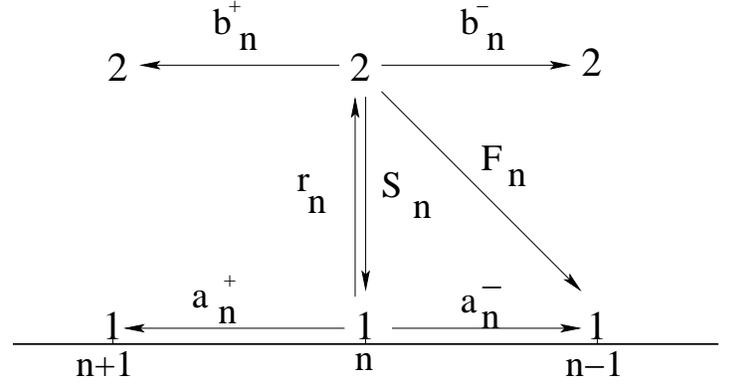}
\end{center}
\caption{A schematic representation of the position of the DNA-histone contact with two state of the model.}
\label{fig-model2}
\end{figure}
\subsection{Footprint traversal time}

We define $P_{\mu,n}(t) = \sum_{m} P_{\mu}(m,n,t)$ as the probability 
that the $n$ histone-DNA contacts are intact at time $t$, irrespective 
of the position of the CRE motor. 
From equations (\ref{eq-mn1})-(\ref{eq-mn10}), summing over $m$, we get 
the following equations: 
For n$\geq$(N+1)
\begin{eqnarray}
dP_{1,n}/dt&=&p_{N}[P_{1,n+1}-2P_{1,n}+P_{1,n-1}]+\nonumber \\
&&\omega^{f}_{12}P_{2,n+1}+\omega_{12}P_{2,n}-\omega_{21}P_{1,n}
\end{eqnarray}
and
\begin{eqnarray}
dP_{2,n}/dt&=&p_{N}[P_{2,n+1}-2P_{2,n}+P_{2,n-1}]+\nonumber \\
&&\omega_{21}P_{1,n}-\omega_{12}P_{2,n}-\omega^{f}_{12}P_{2,n} 
\end{eqnarray}
For n=N
\begin{eqnarray}
dP_{1,N}/dt&=&-p_{N}P_{1,N}+\omega_{12}P_{2,N}+\omega^{f}_{12}P_{2,N+1}\nonumber\\
&&+g_{0}p_{N-1}P_{1,N-1}+g_{1}p_{N}P_{1,N}+p_{N}\nonumber\\
&&P_{1,N+1}+f_{1}p_{N}P_{2,N}-\omega_{21}P_{1,N}
\end{eqnarray}
and
\begin{eqnarray}
dP_{2,N}/dt&=&-(\tilde\omega^{f}_{12}+p_{N})P_{2,N}+\omega_{21}P_{1,N}-\omega_{12}P_{2,N}\nonumber \\
&&+p_{N}P_{2,N+1}+f_{0}p_{N-1}P_{2,N-1} 
\end{eqnarray}
For 3$\leq$ n $\leq$ N
\begin{eqnarray}
dP_{1,n}/dt&=&-p_{n}P_{1,n}+\tilde\omega^{f}_{12}P_{2,n+1}\nonumber \\
&&+\omega_{12}P_{2,n}-\omega_{21}P_{1,n}+g_{0}p_{n-1}P_{1,n-1}+\nonumber \\
&&g_{1}p_{n}P_{1,n}+g_{2}p_{n+1}P_{1,n+1}+f_{1}p_{n}P_{2,n}+\nonumber \\
&&f_{2}p_{n+1}P_{2,n+1} 
\end{eqnarray}
and
\begin{eqnarray}
dP_{2,n}/dt&=&-(\tilde\omega^{f}_{12}+p_{N})P_{2,n}+\omega_{21}P_{1,n}\nonumber \\
&&-\omega_{12}P_{2,n}+f_{0}p_{n-1}P_{2,n-1} 
\end{eqnarray}
For n$=$1
\begin{eqnarray}
dP_{1,1}/dt&=&-k_{u}P_{1,1}+\tilde\omega^{f}_{12}P_{2,2}+\nonumber \\
&&f_{2}p_{2}P_{2,2}+g_{2}p_{2}P_{1,2}+\omega_{12}P_{2,1}-\nonumber \\
&&\omega_{21}P_{1,1}
\end{eqnarray}
and
\begin{eqnarray}
dP_{2,1}/dt=-k_{u}P_{2,1}+\omega_{21}P_{1,1}-\omega_{12}P_{2,1}
\end{eqnarray}
For n$=$2
\begin{eqnarray}
dP_{1,2}/dt&=&-p_{2}P_{1,2}+\tilde\omega^{f}_{12}P_{2,3}\nonumber \\
&&+g_{1}p_{2}P_{1,2}+f_{1}p_{2}P_{2,2}+g_{2}p_{3}P_{1,3}\nonumber \\
&&+f_{2}p_{3}P_{2,3}+\omega_{12}P_{2,2}-\omega_{21}P_{1,2} 
\end{eqnarray}
and
\begin{eqnarray}
dP_{2,2}/dt&=&-\omega_{12}P_{2,2}+\omega_{21}P_{1,2}-p_{2}P_{2,2}\nonumber \\
&&-\tilde\omega^{f}_{12}P_{2,2} 
\end{eqnarray}

We define the {\it survival probability} $S_{\mu,n}(t)$ to be the 
probability that the CRE has not yet reached the far end of the 
footprint till time $t$, given that initially (at $t=0$) there 
were $n$ intact contacts between the histone spool and the DNA 
on the footprint in front of the CRE motor. Obviously, 
$S_{\mu,n}(t)$ is the solution of the equations for $P_{\mu,n}(t)$ 
with the initial condition $S_{\mu,n}(0)=1$.

Interestingly, the time-evolution of $S_{\mu,n}(t)$ can be re-cast as  
\begin{eqnarray}
dS_{1,n}/dt&=&a_{n}^{+}(S_{1,n+1}-S_{1,n})+a_{n}^{-}(S_{1,n-1}-S_{1,n})\nonumber \\
&&+r_{n}(S_{2,n}-S_{1,n})
\label{eq-s1}
\end{eqnarray}
\begin{eqnarray}
dS_{2,n}/dt&=&b_{n}^{+}(S_{2,n+1}-S_{2,n})+b_{n}^{-}(S_{2,n-1}-S_{2,n})\nonumber \\
&&+s_{n}(S_{1,n}-S_{2,n})+F_{n}(S_{1,n-1}-S_{2,n})
\label{eq-s2}
\end{eqnarray}
where the transition rates $a^{\pm}_{n},b^{\pm}_{n},r_{n},s_{n}$ and $F_{n}$ depend 
on the value of $n$ as follows: 
\begin{widetext}
For $n\geq(N+1)$\newline \\
$F_{n}=\omega^{f}_{12},a_{n}^{+}=p_{N},a_{n}^{-}=p_{N},b_{n}^{+}=p_{N},b_{n}^{-}=p_{N},r_{n}=\omega_{21},s_{n}=\omega_{12}$\newline \\
For n=N \newline \\
$F_{n}=\omega^{f}_{12},a_{n}^{+}=g_{0}p_{N},a_{n}^{-}=p_{N},b_{n}^{+}=f_{0}p_{N},b_{n}^{-}=p_{N},r_{n}=\omega_{21},s_{n}=(\omega_{12}+f_{1}p_{N})$ \newline \\
For $3\leq n < N$ \newline \\
$F_{n}=(\tilde\omega^{f}_{12}+f_{2}p_{n}),a_{n}^{+}=g_{0}p_{n},a_{n}^{-}=g_{2}p_{n},b_{n}^{+}=f_{0}p_{n},b_{n}^{-}=0,r_{n}=\omega_{21},s_{n}=(\omega_{12}+f_{1}p_{n})$ \newline \\
For $n=2$ \newline \\
$F_{2}=\tilde\omega^f_{12}+f_{2}p_{2},a_{2}^{+}=0,a_{2}^{-}=g_{2}p_{2},b_{2}^{+}=0,b_{2}^{-}=0,r_{2}=\omega_{21},s_{2}=\omega_{12}+f_{1}p_{2}$ \newline \\
For $n=1$ \newline \\
$F_{1}=0,a_{1}^{+}=0,a_{1}^{-}=k_{u},b_{1}^{+}=0,b_{1}^{-}=k_{u},r_{1}=\omega_{21},s_{1}=\omega_{12}$\newline \\
\end{widetext}
The master equations (\ref{eq-s1})-(\ref{eq-s2}) together, effectively, correspond 
to the kinetic scheme shown in the Fig. \ref{fig-model2}. Using this scheme, the 
MFTT for the single CRE motor can be calculated analytically 
by extending the theoretical framework developed in ref.\cite{pury03} for calculating 
the mean first-passage time of random walks.  

Following Pury and Caceres \cite{pury03}, the MFTT is given by 
\begin{equation}
T_{\mu,n}=\int_{0}^{\infty}S_{\mu,n}(t)dt 
\end{equation}
Since $S_{\mu,n}(\infty) = 0$ and $S_{\mu,n}(0)=1$, integrating the equations 
(\ref{eq-s1}) and (\ref{eq-s2}) with respect to $t$, we get 
\begin{eqnarray}
-1&=&a^+_n [T_{1, n+1} - T_{1,n}] + a^-_{n}[T_{1,n-1}-T_{1,n}] \nonumber \\ 
&+&r_n[T_{2,n} - T_{1,n}]
\label{eq-T1}
\end{eqnarray}
\begin{eqnarray}
-1 &=& b^+_n(T_{2,n+1} - T_{2,n}) + b^-_n(T_{2,n-1}-T_{2,n}) \nonumber \\
&+&s_n(T_{1,n}-T_{2,n}) + F_n(T_{1,n-1} - T_{2,n})
\label{eq-T2}
\end{eqnarray}

Defining 
\begin{eqnarray} 
\Delta_{\mu,n} &=& T_{\mu,n+1} - T_{\mu,n}, \nonumber \\ 
\delta_{n} &=& T_{2,n} - T_{1,n} 
\end{eqnarray} 
equations (\ref{eq-T1}) and (\ref{eq-T2}) can be expressed as 
\begin{equation}
-1 = a_n^+ \Delta_{1,n} - a^-_n \Delta_{1,n-1} + r_n \delta_n 
\label{eq-T3}
\end{equation}
\begin{equation} 
-1 = b^+_n \Delta_{2,n} - b^-_n \Delta_{2,n-1} - s_n\delta_n -F_n(\Delta_{1,n-1}+\delta_n)
\label{eq-T4}
\end{equation}

Now, in the special case 
\begin{equation}
\Delta_{1,n} = \Delta_{2,n} = \Delta_n, 
\label{eq-5a}
\end{equation}
equations (\ref{eq-T3}) and (\ref{eq-T4}) become 
\begin{equation}
-1 = a_n^+ \Delta_{n} - a^-_n \Delta_{n-1} + r_n \delta_n 
\label{eq-T6}
\end{equation}
\begin{equation}
-1 = b^+_n \Delta_{n} - (b^-_n+F_n) \Delta_{n-1} -( s_n+F_n)\delta_n 
\label{eq-T7}
\end{equation}
Next, multiplying equation (\ref{eq-T6}) by $(s_n+F_n)$ and Eq. 
(\ref{eq-T7}) by $r_n$, and then adding the resulting equations, 
we get 
\begin{eqnarray}
-(r_n+s_n+F_n) &=& \{r_n b_n^+ + (s_n+F_n)a^+_n \}\Delta_n \nonumber \\
&-& \{  r_n(b^-_n + F_n) + (s_n+F_n)a^-_n  \}\Delta_{n-1} \nonumber \\
\label{eq-T8}
\end{eqnarray}
Eq. (\ref{eq-T8}) can be re-written as 
\begin{equation}
-C_n = B_n\Delta_n - A_n\Delta_{n-1}
\label{eq-T9}
\end{equation}
where, 
\begin{eqnarray} 
C_n &=& (r_n+s_n+F_n), \nonumber \\
B_n &=& \{r_n b_n^+ + (s_n+F_n)a^+_n \}, \nonumber \\
A_n &=& \{  r_n(b^-_n + F_n) + (s_n+F_n)a^-_n  \} 
\end{eqnarray}
We can rewrite Eq. (\ref{eq-T9}) as follows
\begin{equation}
 \Delta_{n-1} = \frac{B_{n}}{A_{n}}\Delta_n +\frac{C_n}{A_n}
\label{eq-T10}
\end{equation} 

For a {\it fully wrapped} histone, the MFTT $t_d$ is given by 
\begin{equation}
t_{d}=\sum_{n=1}^{N}\triangle_{n} 
\label{eq-T11}
\end{equation}
Using (\ref{eq-T10}) in (\ref{eq-T11}) we, finally, get
\begin{eqnarray}
t_d=\sum_{n=1}^{N}[C_n/A_n+\sum_{i=1}^{\infty}C_{n+i}/A_{n+i}\prod_{k=0}^{i-1}B_{n+k}/A_{n+k}]
\label{eq-T12}
\end{eqnarray}

Since it is not easy to get an intuitive feeling for the implications 
of the expression (\ref{eq-T12}), we anaylyze its special simpler 
forms in some limiting cases. 
In the limit of extremely slow motor, i.e., $\omega^{f}_{12} \to 0$, 
as expected, the expression (\ref{eq-T12}) for the MFTT 
$t_d$ {\it diverges}.

\begin{figure}
 \begin{center}
  \includegraphics[width=0.45\columnwidth]{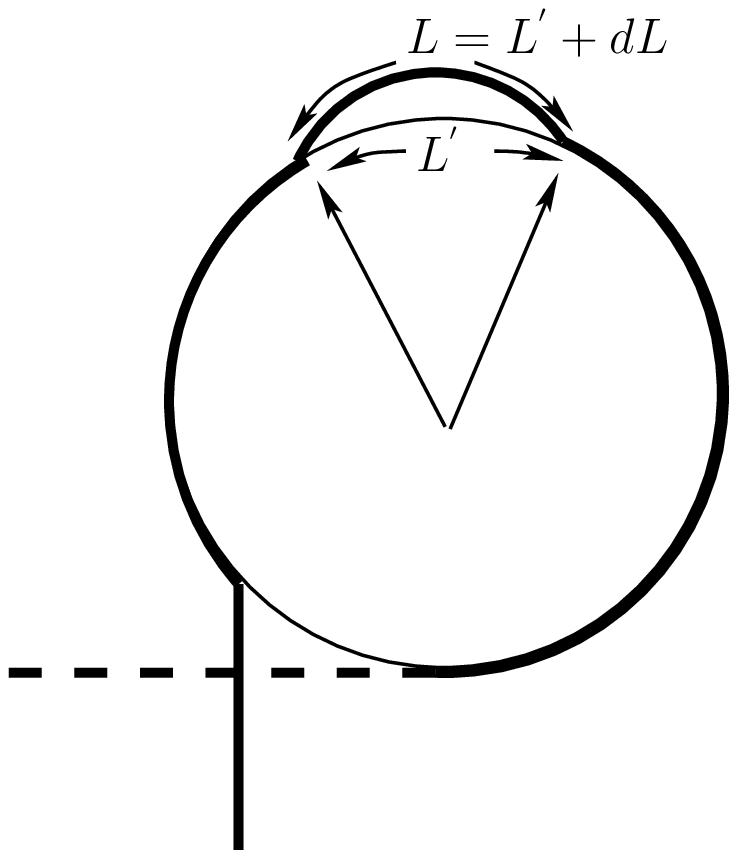}
\caption{Top view of the histone octamer bound with DNA. Loop formation involving length $dL$ of linker chain being incorporated
into the nucleosome, with length $L'$ of the exposed surface. (adapted from Fig. 2 of ref. \cite{schiessel01}).}
\label{fig}
 \end{center}
\end{figure}

For ensuring high-speed of the CRE motor, we need simultaneously 
$w^{f}_{12}/k_b >>1$ and $w_{21}/k_b >>1$. If, for simplicity, we 
make the additional assumption that $w^{f}_{12}$ is the slower of 
the two, i.e., $\omega_{21} >> \omega^f_{12}$, we have 
$f_0=f_1=g_0=g_1\simeq0$ and $f_2 \simeq 1$, and $g_2 \simeq 1$. 
Hence, in this limit, $B_{n} = 0$ for $n \leq N$ and, therefore, 
\begin{equation}
t_d = \sum_{n=1}^{N}\frac{C_n}{A_n} \simeq N/k_{u} 
\label{eq-liminfty}
\end{equation}
which is identical to the corresponding limiting value of $t_d$ 
reported in ref.\cite{chou}. This is a consequence of the fact that 
in the limit of extremely fast motor, because of the assumption of 
very large value of $\omega_{21}$, the 2-state model reduces to an 
effectively 1-state model.
We make a numerical estimate of $t_d$ in this limit by computing an 
approximate value of $k_{u}$. Defining  
\begin{equation}
K=k_b/k_u
\label{eq-K}
\end{equation}
as the flap binding constant, we can rewrite the equation 
(\ref{eq-liminfty}) as 
\begin{equation}
t_d = \frac{NK}{k_b}
\label{eq-liminfty2}
\end{equation}
Range of typical values of $K$ has been 
used earlier by Chou \cite{chou}. Using this range of values for 
$K$, one can estimate $k_u$, provided a typical value of $k_b$ is known.
Therefore, we now estimate the typical numerical values of $k_b$ 
following Schiessel and coworkers \cite{schiessel01,kulic03a,kulic03b}.  
Suppose, $L_0 (\simeq 50$ nm) be the length of the DNA that wraps around 
the histone spool. Let $L^{'}+dL$ be the contour length of the loop 
induced by spontaneous thermal fluctuations where (see Fig. \ref{fig}) 
$L^{'}$ is the exposed arc length on the histone spool that was 
covered by the DNA segment prior to the loop formation and $dL$ is a 
small segment of the linker dsDNA that has been pulled into the loop.

We assume that the life time of a loop $(\tau)$ is much shorter 
than the average time required to form a loop. Following Schiessel et al. 
\cite{schiessel01,kulic03a,kulic03b} we write down the rate of loop 
formation as
\begin{equation}
\alpha \simeq \frac{L_0 exp(-E_{bend}/(k_BT))}{\tau L^{'}}
\label{eq-al_est2} 
\end{equation} 
Comparing Eqs. (\ref{eq-al_est1}) and (\ref{eq-al_est2}) we obtain
\begin{equation}
k_b = \frac{L_0}{\tau L^{'}}.
\label{eq-kb-est1}
\end{equation}
Since $\tau^{-1}=k$ characterizes the rate of unbiased diffusion of the 
loop around the histone spool \cite{schiessel01}
$\tau \simeq \frac{L_0^2}{D}$
where $D$ is the corresponding diffusion constant. From Stokes-Einstein 
relation $D = k_BT/\zeta$, where $\zeta \simeq \eta L^{'}$ \cite{schiessel01} 
and $\eta$ is the effective viscosity of the aqueous medium. Combining all 
the results and substituting these into Eq. (\ref{eq-kb-est1}) we finally 
obtain \begin{equation}
k_b = \frac{k_BT}{\eta L_0 (L^{'})^2}
\label{eq-kb-est2}
\end{equation} 
The estimation can be completed only if an estimate of $L^{'}$ is available. 
Following ref. \cite{schiessel01} (Eq. (2a) of \cite{schiessel01}), we get  
\begin{equation}
L^{'} \simeq (\frac{20\pi^4\kappa}{\lambda R_0^2})^{1/6}(dL/R_0)^{1/3}R_0,
\label{eq-Lp}
\end{equation}
where $\kappa$ is the bending elastic constant of the semi-flexiable DNA 
chain, $\lambda$ is the adsorption energy per unit length and $R_0$ is 
the radius of the histone spool. Using the reasonable values quoted in 
ref.\cite{schiessel01,kulic03a,kulic03b}, namely, 
$R_0 = 5 nm, \kappa = 207.10 pN-nm^2, dL=3.40 nm$ and $\lambda = 5.92 pN$ 
we obtain from Eq. (\ref{eq-Lp}) $L^{'}=16.43 nm$. using this estimate 
of $L^{'}$, together with  $1k_BT = 4.142$ pN-nm, $L_0 = 500\AA$, and 
$\eta = 1$ Centipoise, we obtain from 
Eq. (\ref{eq-kb-est2}) $k_b = 306877.4 s^{-1}$. 

With the above estimated value of $k_b,$ and $N=15$ from 
Eq. (\ref{eq-liminfty2}) we get the estimates $t_d = 0.000024 s$ 
for $K=0.5$ and $t_d = 0.0005 s$ for $K=10$. 
Such small values of $t_d$, estimated from Eq.\ref{eq-liminfty}, 
arise from the fact that the approximate form (\ref{eq-liminfty}) is 
valid only in the limit of extremely fast motor. Therefore, this 
limiting formula provides only a lower bound and does not correspond 
to real CRE motors under physiological conditions.

\begin{figure}
\begin{center}
\includegraphics[scale=0.7,angle=0]{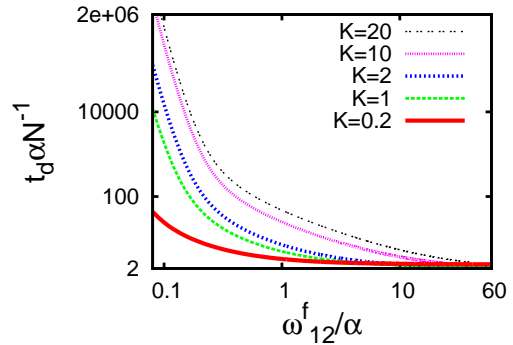}
\caption{Normalized MFTT $\alpha N^{-1} t_{d}$ plotted 
against the normalized motor speed $\omega ^{f}_{12}/\alpha$ for 
different values of $K$ with $\omega_{12}/\alpha=\omega_{21}/\alpha=0.5$.}
\label{fig-survtm}
\end{center}
\end{figure}
In Fig.{\ref{fig-survtm}} we plot the normalized MFTT 
$t_d \alpha/N$ as a function of the normalized motor speed 
$\omega^{f}_{12}/\alpha$ 
for $\omega_{12}/\alpha=\omega_{21}/\alpha=0.5$ and a few  
fixed values of the parameter $K$. For any fixed value of $K$, 
the normalized MFTT decreases monotonically with the 
increase of the normalized motor speed and saturates to the value 
given by equation (\ref{eq-liminfty}) in the limit 
$\omega^{f}_{12}/\alpha \to \infty$. Moreover, for a given value 
of $\omega^{f}_{12}/\alpha$, as the flap binding constant $K$ 
increases the MFTT increases.

\begin{figure}
\begin{center}
(a) \\
\vspace*{.5cm}
\includegraphics[scale=0.7,angle=0]{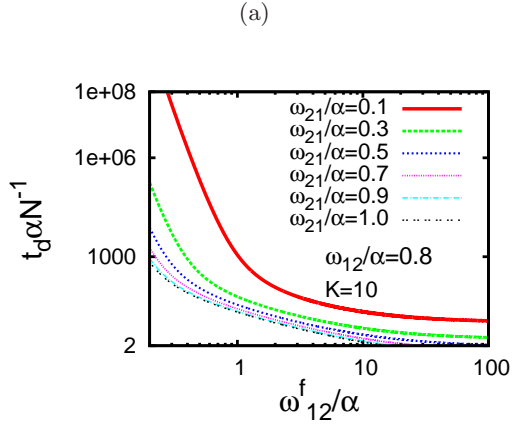}\\
(b) \\
\vspace*{.5cm}
\includegraphics[scale=0.7,angle=0]{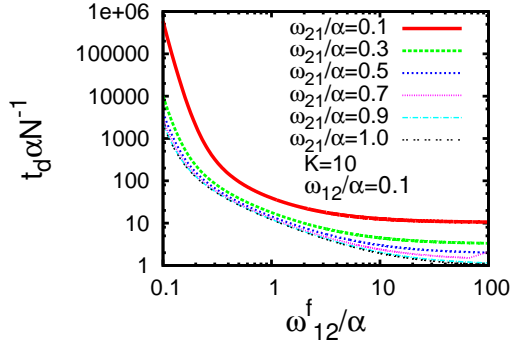}
\caption{Normalized MFTT $\alpha N^{-1} t_{d}$ plotted 
against the normalized motor speed $\omega ^{f}_{12}/\alpha$ for 
different values of $\omega_{21}/\alpha$ with (a) $\omega_{12}/\alpha=0.8$, 
and (b) $\omega_{12}/\alpha=0.1,K=10$.}
\label{fig-survtm-hw12}
\end{center}
\end{figure}

In Fig.{\ref{fig-survtm-hw12}}, $t_d$ is plotted against 
$\omega^{f}_{12}/\alpha$ for 
(a) $K=10$, $\omega_{12}/\alpha=0.8$, and 
(b) $K=10$, $\omega_{12}/\alpha=0.1$, each  
for a few distinct values of $\omega_{21}/\alpha$. 
The MFTT decreases as $\omega_{21}/\alpha$ increases. 
This is a consequence of the fact that $\omega_{21}$ depends on 
the ATP concentration. For small $\omega_{12}/\alpha$ reduces the amplitude of peeling time. 

\begin{figure}
\begin{center}
(a) \\
\vspace*{.5cm}
\includegraphics[scale=0.7,angle=0]{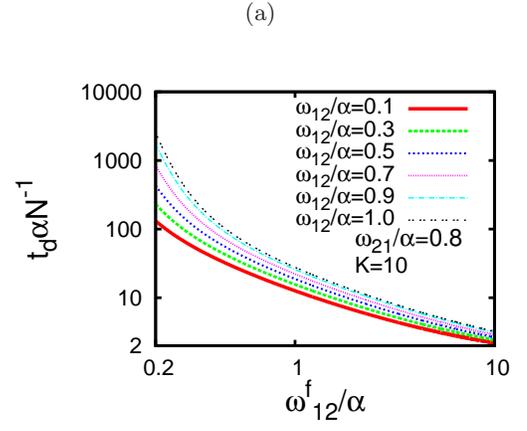}\\
(b) \\
\vspace*{.5cm}
\includegraphics[scale=0.7,angle=0]{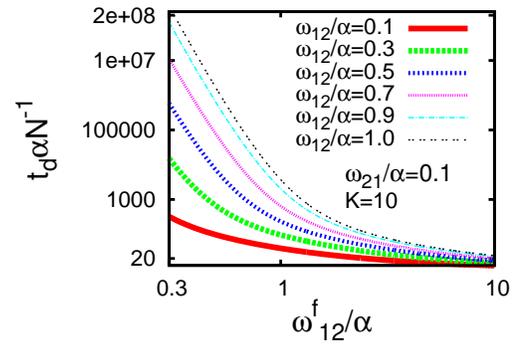}
\caption{Normalized MFTT $\alpha N^{-1} t_{d}$ plotted 
against the normalized motor speed $\omega ^{f}_{12}/\alpha$ for 
different values of $\omega_{12}/\alpha$ with (a) $\omega_{21}/\alpha=0.8$, 
and (b) $\omega_{21}/\alpha=0.1,K=10$.}
\label{fig-survtm-hw21}
\end{center}
\end{figure}

In Fig.{\ref{fig-survtm-hw21}} we demonstrate that for large value 
of $\omega_{21}/\alpha$, which effectively speeds up the motor, 
reduces the magnitude of the MFTT. 

Although the qualitative trends of variations of $t_d$ with 
$\omega^{f}_{12}/\alpha$ in our model is similar to that in 
Chou's model \cite{chou}, wide range of variation of $t_d$ 
is possible in our model by controlling $\omega_{21}$ which, 
in turn, can be controlled by the ATP concentration.

\begin{figure}
\begin{center}
\includegraphics[scale=0.7,angle=0]{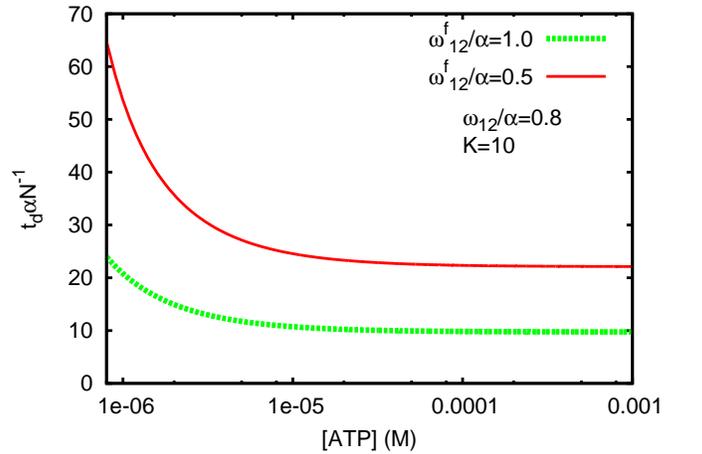}
\caption{Normalized MFTT $\alpha N^{-1} t_{d}$ plotted 
against the ATP concentration [ATP] for $\omega_{12}/\alpha=0.1$, 
$K=10.0$ and two different values of $\omega^{f}_{12}/\alpha$.} 
\label{fig-survtm-ATP1}
\end{center}
\end{figure}

In order to explore the dependence of $t_d$ on the concentration of 
ATP, we first assume that 
\begin{equation}
\omega_{21} = \omega_{21}^{0} [ATP] 
\label{eq-ATP1} 
\end{equation}
Assuming a typical value $\omega_{21}^{0} = 10^6 M^{-1} s^{-1}$, 
we have plotted the normalized MFTT against the ATP 
concentration for two different normalized values of the 
unhindreed motor speed keeping the other parameters fixed. 
With the increase of ATP concentration, the MFTT decreases
and, gradulally saturates. When ATP concentration is sufficiently 
high, the step with rate constant $\omega_{21}$ is no longer 
rate-limiting. We also find that, for a given ATP concentration, 
the higher is the value of $\omega^f_{12}/\alpha$ the shorter is 
the MFTT $t_d$. 

The linear dependence of $\omega_{21}$ on ATP concentration, as 
envisaged in (\ref{eq-ATP1}), may be valid only at sufficiently low 
concentration of ATP. In general, $\omega_{21}$ may follow the 
usual {\it Michaelis-Menten equation} for the rate of enzymatic 
reactions (because $\omega_{21}$ represents the rate of ATP hydrolysis 
catalyzed by the CRE motor) \cite{dixon79}. In that case $\omega_{21}$ 
itself would saturate with the increase of ATP concentration, instead 
of increasing linearly with [ATP].

\section{Conclusion}

In this paper we have studied the process of ATP-dependent chromatin 
remodeling. For simplicity, we have considered only a single nucleosome 
consisting of a dsDNA strand wrapped one and three-fourth turns around 
a cylindrical spool made of histone proteins. We have extended Chou's 
model \cite{chou} by assigning two distinct ``chemical'' states to the 
CRE and postulating a minimal mechano-chemical kinetic scheme for 
capturing the effects of ATP hydrolysis explicitly. Our theoretical 
framework has been developed exploiting a close analogy with the 
unzipping of a double-stranded DNA by a helicase \cite{garai}. 
We have written down the master equations for the postulated kinetic 
scheme. This model of footprint traversal by ATP-dependent CRE can be 
easily interpreted as an implementation of a Brownian ratchet mechanism. 
From an analytical treatment of this stochastic kinetic model, we have 
derived analytical expression for the MFTT of  
the ATP-dependent CRE. We make explicit analytical 
predictions on the dependence of the MFTT on (i) the 
unhindred speed of the CRE, as well as on (ii) the concentration of ATP. 
In principle our theoretical predictions can be tested by carrying out 
{\it in-vitro} experiments with a single nucleosome.


{\bf Acknowledgements:} 
DC thanks Michael Poirier and Tom Chou for useful discussion and  
correspondence, respectively. 
DC acknowledges support of the Visitors Program of the Max-Planck 
Institute for the Physics of Complex Systems in Dresden where parts of 
this work were carried out during two separate visits. DC also thanks 
Frank J\"ulicher for discussions in the initial stages of this work 
and Anirban Sain for a critical reading of the manuscript.
This work is also supported, in part, by IIT Kanpur through the Dr. Jag 
Mohan Chair professorship (DC) and by a research grant from CSIR, India (DC). 
AG thanks UGC, India, for a senior research fellowship. 



\begin{thebibliography}{99}
\bibitem{wolffe98} A. Wolfee, {\it Chromatin: structure and function}, 
(Academic Press, 1998). 
\bibitem{widom98} J. Widom, Annu. Rev. Biophys. Biomol.Struc. {\bf 27}, 
285 (1998).
\bibitem{schiessel03} H.Schiessel, J.  Phys. Condens. Matter, {\bf 15}, 
R699 (2003).
\bibitem{lavelle06} C. Lavelle and A. Benecke, Eur. Phys. J. E {\bf 19}, 
379 (2006).
\bibitem{kornberg99} R.D. Kornberg and Y. Lorch, Cell {\bf 98}, 285 
(1999).
\bibitem{polach95} K.J. Polach and J. Widom, J. Mol. Biol. {\bf 254}, 
130 (1995).
\bibitem{anderson02} J.D. Anderson, A. Thastrom and J. Widom,
Mol. Cell. Biol. {\bf 22}, 7147 (2002).
\bibitem{fan03} H.Y. Fan, X. He, R.E. Kingston and G.J. Narlikar, 
Mol. Cell {\bf 11}, 1311 (2003).


\bibitem{flaus03} A. Flaus and T. Owen-Hughes, Biopolymers {\bf 68}, 
563 (2003). 
\bibitem{flaus11} A. Flaus and T. Owen-Hughes, FEBS J. {\bf 278}, 
3579 (2011).
\bibitem{halford04a} S.E. Halford, A.J. Welsh and M.D. Szczelkun,
Annu. Rev. Biophys. Biomol. Struct.  {\bf 33}, 1 (2004).
\bibitem{saha06} A. Saha, J. Wittmeyer and B.R. Cairns, 
Nat. Rev. Mol. Cell Biol. {\bf 7}, 437 (2006). 
\bibitem{clapier09} C.R. Clapier and B.R. Cairns, Annu. Rev. Biochem. 
{\bf 78}, 273 (2009). 
\bibitem{narlikar08} L. R. Racki and G. J. Narlikar, Curr. Opin, 
Genet. Dev. {\bf 18}, 137 (2008).

\bibitem{sakaue01} T. Sakaue, K. Yoshikawa, S.H. Yoshimura and
K. Takeyasu, Phys. Rev. Lett. {\bf 87}, 078105 (2001).

\bibitem{schiessel01} H. Schiessel, J. Widom, R.F. Bruinsma and 
W.M. Gelbert, Phys. Rev. Lett. {\bf 86}, 4414 (2001).
\bibitem{kulic03a} I.M. Kulic and H. Schiessel, Phys. Rev. Lett. 
{\bf 91}, 148103 (2003). 
\bibitem{kulic03b} I.M. Kulic and H. Schiessel, Biophys. J. 
{\bf 84}, 3197 (2003).
\bibitem{schiessel06} H. Schiessel, Eur. Phys. J. E {\bf 19}, 
251 (2006).
\bibitem{rafiee04} F. Mohammad-Rafiee, I.M. Kulic and H. Schiessel,
J. Mol. Biol. {\bf 344}, 47 (2004). 
\bibitem{blossey11} R. Blossey and H. Schiwssel, FEBS J. {\bf 278}, 
3619 (2011). 
\bibitem{mobius06} W. M\"obius, R.A. Neher and U. Gerland, 
Phys. Rev.  Lett. {\bf 97}, 208102 (2006).
\bibitem{langowski06} J. Langowski, Eur. Phys. J. E {\bf 19}, 241 
(2006).
\bibitem{lense06} A. Lense and J.M. Victor, Eur. Phys. J. E
{\bf 19}, 279 (2006).
\bibitem{vaillant06} C. Vaillant, B. Audit, C. Thermes and A. Arneodo,
Eur. Phys. J. E {\bf 19}, 263 (2006).
\bibitem{ranjith07} P. Ranjith, J. Yan and J.F. Marko, PNAS {\bf 104}, 
13649 (2007).
\bibitem{lia06} G. Lia, E. Praly, H. Ferreira, C. Stockdale,
Y.C. Tse-Dinh, D. Dunlap, V. Croquette, D. Bensimon and T. Owen-Hughes,
Mol. Cell {\bf 21}, 417 (2006).
\bibitem{cairns07} B.R. Cairns, Nat. Struct. and Mol. Biol. {\bf 14}, 
989 (2007).

\bibitem{chou} T. Chou, Phys. Rev. Lett., {\bf 99}, 058105 (2007).

\bibitem{li04} G. Li, M. Levitus, C. Bustamante and J. Widom, {\it 
Rapid spontaneous accessibility of nucleosomal DNA}, Nat. Struct. 
Mol. Biol. {\bf 12}, 46-53 (2004). 
\bibitem{poirier08} M.G. Poirier, M. Bussiek, J. Langowski, J. Mol. 
Biol. {\bf 379}, 772-786 (2008). 
\bibitem{poirier09} M.G. Poirier, E. Oh, H.S. Tims and J. Widom, 
Nat. Struct. Mol. Biol. {\bf 16}, 938 (2009). 
\bibitem{tims11} H.S. Tims, K. Gurunathan, M. Levitus and J. Widom, 
J. Mol. Biol. {\bf 411}, 430 (2011).

\bibitem{meerssman92} G. Meersseman, S. Pennings and E.M. Bradbury, 
EMBO J. {\bf 11}, 2951 (1992).  

\bibitem{lusser03} A. Lusser and J.T. kadonaga, BioEssays {\bf
25}, 1192 (2003).
\bibitem{becker02a} P.B. Becker and W. H\"orz, Annu Rev. Biochem. 
{\bf 71}, 247-273 (2002).
\bibitem{langst04} G. L\"angst and P.B. Becker, Biochim. Biophys. Acta 
{\bf 1677}, 58 (2004).

\bibitem{garai} A. Garai, D. Chowdhury, M. Betterton, {\bf 77}, 061910 
(2008).

\bibitem{pury03} P.A. Pury and M. O. Caceres, J. Phys. A {\bf 36}, 
2695 (2003).
\bibitem{dixon79} M. Dixon and E.C. Webb, {\it Enzymes} (Academic Press, 1979).

\end{thebibliography}
\end{document}